\documentclass[structabstract]{aa}
\usepackage{graphicx}

\usepackage{txfonts}
\begin{document}
   \title{The activity of Main Belt comets}

    \author{M. T. Capria
          \inst{1}
          \and
          S. Marchi\inst{2}
          \and
          M. C. De Sanctis\inst{1}
          \and
          A. Coradini\inst{3}
          \and
          E. Ammannito\inst{3}
          }

   \institute{INAF - IASF, Area Ricerca Tor Vergata,
              Via del Fosso del Cavaliere 100, 00133 Rome\\
              \email{mariateresa.capria@iasf-roma.inaf.it}
         \and
             Departement Cassiop\'ee, Universit\'e de Nice - Sophia Anthipolis,\\
                  Observatoire de la C\^ote d'Azur, CNRS, boulevard de l'Observatoire, B.P. 4299,\\
                  0634 Nice Cedex 4, France
             \and
             INAF - IFSI, Area Ricerca Tor Vergata,
              Via del Fosso del Cavaliere 100, 00133 Rome\\
             }

   \date{Received September 15, 1996; accepted March 16, 1997}


  \abstract
   {Main Belt comets represent a recently discovered class of objects. They are quite intriguing because, while having a
Tisserand invariant value higher than 3, are showing cometary activity.  }
   {We study the activity of the Main Belt comets making the assumption that they are icy-bodies and that the activity
has been triggered by an impact. We try to determine the characteristics of this activity. We also try to determine if
the  nowadays impact rate in the Main Asteroid Belt is compatible with the hypothesis of an activity triggered by a
 recent impact. }
   {Due to the fact that the Main Belt comets can be considered as a kind of comets, we apply a thermal evolution
 model developed for icy bodies in order to simulate their activity. We also apply a model to derive the impact rate, with respect to the size of the impactor,
in the Main Belt. }
   {We demonstrate that a stable activity can result from a recent impact, able to expose ice-rich layers, and that
the impact rate in the Main Belt is compatible with this explanation.  }
   {}


   \maketitle
%

\section{Main Belt comets: a new class of objects}

The Main Belt Comets (MBCs) are a recently discovered class of objects, composed until now by only four objects
(Table~\ref{tab1}). These objects are puzzling because they are orbiting in the Main Belt on stable orbits, with
Tisserand invariant with respect to Jupiter higher than 3, but at the same time they are showing a more or less
stable cometary activity. Four known objects implies many more presently inactive or faintly active and as a
consequence undetectable. It is probable that as a result of the surveys being conducted
(Gilbert \& Wigert \cite{gilbert}) more objects will be added to this new class in a near future.  \\
The first of the MBCs that has been discovered is 133P/Elst-Pizarro. At the moment of the discovery,  in 1992
(Elst et al. \cite{elst}), it has been designated as 7968 Elst-Pizarro. When in July 1996 the body was found showing a
dust tail, the activity was initially attributed to an impact (Boehnhardt et al. \cite{boehnhardt}; Toth \cite{toth}),
 but when later on the object was found periodically active at the successive perihelions (2002 and then 2007) it
became clear that a different explanation had to be found (Hsieh et al. \cite{hsieh0}). P/2005 U1 (Read) and
176P/LINEAR followed in a short time (Read et al. \cite{read}; Green \cite{green}), both active and orbiting on similar
orbits (Table~\ref{tab1}). More recently, P/2008 R1 (Garradd) has been discovered (Garradd et al. \cite{garradd}).  \\
The Tisserand invariant values rule out the possibility that these bodies are newcomers in this part of the Solar
System. The orbits of two of them, 133P/Elst-Pizarro and 176P/LINEAR, can be linked with the 2 Gyr-old Themis
dynamical family (Zappal\'a et al. \cite{zappala}) and the smaller and younger (10 Myr) Beagle family
(Nesvorny et al. \cite{nesvorny}). The orbit of P/2005 U1 (Read) is falling just outside the Themis family, and
P/2008 R1 (Garradd) appears to be a dynamically unstable object, probably formed in a different part of the Main Belt
(Jewitt et al. \cite{jewitt1}). \\
As for the causes of dust ejection, it became soon evident that it is due to sublimation. Not only this activity is
recurrent, as in the case of 133P/Elst-Pizarro, but it is also lasting for weeks and months. Moreover, dust velocities,
estimated from the sunward extent of the coma, are greatly in excess of what would be expected from electrostatic
levitation or rotational ejection (Hsieh \& Jewitt \cite{hsieh}). \\
If these bodies formed in-situ, the question of how can they still contain ice arises. In the protoplanetary disk,
opacity due to grains inhibited radiative cooling and raised the mid-plane kinetic temperature
(Jewitt et al. \cite{jewitt}). Growth and migration of solids in the disk would have caused the opacity to change, so
moving the snow line, that for a fraction of a million years may have been within the Main Belt. In this way, asteroids
 orbiting in the Main Belt could have incorporated water ice upon formation. It is well know that many asteroids lack
the hydration feature in their spectra. This has been interpreted as a sign that their ice never reached the liquid
phase and so could still be found frozen in the interior (Jones et al. \cite{jones}; Carvano et al. \cite{carvano}).
These bodies would be possible candidates for the new class of MBCs.\\
The recent detection of water ice on the surface of 24 Themis (Campins et al. \cite{campins};
Rivkin \& Emery \cite{rivkin}) came as no surprise, and give a strong support to the sublimation-driven explanation for
 the MBCs activity. At the end of 2009, two independent teams observing 24 Themis in the infrared range detected the
presence of water ice widespread on the surface of the body. It has been interpreted as frost, replenished by some
subsurface source. This detection, and those of Main Belt comets, are contributing to definitively shatter the old,
traditional view of the Main Belt as the realm of rocky bodies as opposed to the icy, active bodies that can be found farther.

   \begin{table*}
     \begin{minipage}[t]{\columnwidth}
     \caption[]{Main Belt Comets (as of May 2011)}
     \label{tab1}
     \centering
     \renewcommand{\footnoterule}{}
     \begin{tabular}{c c c c c c c}
     \hline
     Name  & a (AU)\footnote{Semimajor axis}  & {\it e}\footnote{Eccentricity}
     & {\it i} ($^{o}$)\footnote{Inclination} & P (year)\footnote{Orbital period} & TJ\footnote{Tisserand
     parameter}    \\
     \hline
     133P/Elst-Pizarro & 3.156 & 0.165 & 1.39  & 5.62 & 3.184 &  \\
     P/2005 U1 (READ) & 3.196  & 0.192 & 0.24 & 5.63 & 3.166 & \\
     176P/LINEAR & 3.165  & 0.253 & 1.27 & 5.71 &  3.153 \\
     P/2008 R1 (Garradd) & 2.726  & 0.342 & 15.9 & 4.43 & 3.216 \\
     \hline
     \end{tabular}
     \end{minipage}
    \end{table*}

\noindent
The implications deriving from the existence of a third source (besides the Oort Cloud and the Edgeworth-Kuiper Belt) of
"cometary" bodies are remarkable. The water contained in the terrestrial oceans is thought to come from impacting icy
bodies, but the few D/H ratios measured until now in comets have double values with respect to the average ratio obtained
from terrestrial water. A third potential source for Earth water, different from long-period and short-period comets,
could solve the problem, if the D/H ratio in MBCs will prove to be more similar to the terrestrial one.

\section{The activity of Main Belt Comets}

Dust ejection due to sublimation implies the existence of both ice and dust particles. This leads us to imagine that
MBCs are comet-like bodies, that means intimate mixtures of ice and dust, and not rubble-pile rocky objects with some
ice filling the interstices. If this assumption is coupled with the further assumption that the activity is driven by
sublimation, then we can simulate and study the activity of MBCs through the thermal modeling codes used for comets,
KBOs and icy bodies in general.\\
Prialnik and Rosenberg (\cite{prialnik}) applied their comet nuclei thermal evolution model to 177P/Elst-Pizarro. In
their paper the authors  demonstrate that deep-buried ice could have survived since the Main Belt formation time, and
that this ice is most probably composed, at least nowadays, of water crystalline ice. It would have been almost
impossible, for ices sublimating at temperatures well under 130 K, to survive even under an insulating mantle.
Schorghofer too (\cite{schorghofer}) in his paper assumes that the ice, in order to survive for such a long time, must
have been buried under an insulating mantle, and gives estimations of the thickness of this dust layer. The same author
argues against the possibility that the ice be mixed with rocky material rather than dusty material. Rocky surfaces
are seldom able to retain ice, due to the larger thermal conductivity and the larger molecular free path of rocks with
respect to dust grains. \\
Being obvious that exposed ice cannot last for a long time at this distance from the Sun , we make the hypothesis that
buried ice has been exposed in recent times. The most probable explanation is that of an impact excavating a crater in a
dust mantle. This hypothesis, and if the present-days impact frequency in the Main Belt is compatible with it, will be
discussed in a following section. \\
Starting from the assumptions stated above, we apply our thermal evolution model to simulate MBCs activity in the
hypothesis that an impact has recently happened and that a crater has been excavated in the dust mantle. The effect of
the impact could have been to directly expose a fresh ice-rich layer or, if the crater is not deep enough, bring an
ice-rich layer closer to the surface, making it reachable by the the diurnal/seasonal heat wave. The aim of the
simulation is to investigate the activity of MBCs in a more general way and under different hypotheses, and study how
long can the activity last, on which parameters it depends, and which are the characteristics of this activity when a
thin mantle is still covering an active layer.

\subsection{The thermal model}

The thermal evolution model that has been applied to simulate the activity of MBCs is  described in this subsection (Capria et al. \cite {capria}; Capria et al. \cite{capria1}; Capria et al. \cite{capria2}; De Sanctis et al.
\cite{desanctis}, \cite{desanctis1}). In the next subsection, the simulation of dust component, dust flux and mantle
structure are explained in greater detail, because these features have been recently improved in this continuously
evolving code.\\
The model is one-dimensional. The spherical nucleus is porous and composed of ices (water and up to two different species,
typically CO and CO$_2$) and a refractory component. Water ice can be initially amorphous, and in this case  a fraction of
gases such as CO can be trapped in the amorphous matrix and released during the transition to crystalline phase.  In this
work, anyway, we will deal only with water ice in its crystalline status. Together with Prialnik and Rosenberg
(\cite{prialnik}), we deem extremely improbable, for ices sublimating at temperatures well under 130 K, to have
survived till now,  even under an insulating mantle.  \\
Energy and mass conservation is expressed by the following system of coupled differential equations, solved for the whole
nucleus:

\begin{equation}
\rho c \frac{\partial T}{\partial t} =
\nabla [K \cdot \nabla T] +  Q_{i}
\end{equation}

\begin{displaymath}
\frac{1}{RT} \frac{\partial P_i}{\partial t} =
\nabla [G_i \cdot \nabla P_i] + Q\prime_{i}
\end{displaymath}

\noindent where $\rho$, $c$ and $K$ are the bulk density, specific heat and thermal conductivity, $Q_{i}$ is the energy
exchanged by the solid matrix in the sublimation and recondensation of the water ice, $R$ is the gas constant, $P_{i}$ the gas partial
 pressure, $G_{i}$ its diffusion coefficient, and $Q\prime_{i}$ is the gas source term due to
sublimation-recondensation processes.\\
Gas diffusion coefficients are computed on the basis of the mean free path of the molecules in the pore system; the model
 accounts for three different diffusion regimes: Knudsen, viscous and a transition one.\\
The temperature on the surface is obtained by a balance between the solar input and the energy re-emitted in the infrared,
 conducted in the interior and used to sublimate ices present on the surface. The following expression is used:

\begin{eqnarray}
\label{bousur2}
\frac{S(1-A_{s})}{R_{h}^{2}} \frac{\cos \theta
}{\pi} = \epsilon \sigma T_{s}^{4} +
K(T_{s}) \left . \frac {dT}{dr} \right | _{r=R_{n}} + \nonumber \\
f_{H_{2}O}^{s} \cdot H(T_{s})_{H_{2}O} \cdot \dot{\varepsilon} (T_{s})_{H_{2}O}
\end{eqnarray}

\noindent where  $S$ is the solar constant, $R_h$ the heliocentric distance of the comet,  $A_s$ the Bond bolometric
albedo of the surface, $\epsilon$ the infrared emissivity of surface, $\sigma$ the Stefan-Boltzmann constant, $K$ the
thermal conductivity, and $\theta$ the cometary latitude.\\
 In the last term of right member of
 Eq.~(\ref{bousur2}), $f_{H_{2}O}^{s}$ represents the surface fraction covered by water ice, $H$ is the latent heat of
sublimation and $\dot{\varepsilon}$ is the sublimation rate. \\
The surface boundary condition used to solve diffusion equation when ice is present on the surface is:

\begin{equation}
\label{psbound}
P = P^{sat} (T)
\end{equation}

\noindent
When no ice is present, surface boundary condition is obtained assuming free sublimation at comet nucleus surface:

\begin{equation}
\label{phbound}
P = 0
\end{equation}

\noindent
When the temperature rises, ices start to sublimate, beginning from the more volatile ones; even if the nucleus is initially
homogeneous a differentiated, layered structure is always obtained, in which the boundary between different
layers is a sublimation front. Surface erosion due to ice sublimation and particles ejection is taken into account.\\
The dust component is represented as size distributions of spherical grains characterized by a density, a specific heat
and a thermal conductivity; in the code we can consider up to two distributions, with a maximum of ten size classes each.
Dust distributions characteristics and their physical properties were chosen to simulate as much as possible Giotto
findings.
  These experiments onboard of the mission demonstrated that in P/Halley's refractory component it is possible to
evidentiate two chemical phases: an organic one with low atomic weight, called CHON, and a Mg-rich silicate one
(Jessberger and Kissel \cite{jessberger}; McDonnell et al. \cite{mcdonnell}; Mumma \cite{mumma}). Measured particles were in the range of
$10^{-16}-10^{-11}$ g. One third of it did not contain organic components, and the remainder was a mixture of the two
phases; CHON component was probably coating silicate cores. Average densities were found to be ~2500 kg m$^{-3}$ for
silicate grains and ~ 1000 kg m$^{-3}$ for CHON-dominated particles.  \\
For these reasons, the dust component is simulated by two distributions: the first distribution is composed of
silicatic particles, and the second one by mixed silicatic/CHON particles. The characteristics attributed to the two
 distributions are collected in table \ref{tab2}, and are explained hereafter.\\
These two phases have different physical properties (density, thermal conductivity, and so on); as a consequence, the
characteristics (thermal conductivity, tensile strength and cohesion) of a crust mainly composed of silicatic particles
would be different from those of a crust with an
abundant component of refractory organic material.\\

\begin{table}
\caption{Properties of the dust distributions} \label{tab2}
\centering
\begin{tabular}{l c c}
\hline
   Property              & Silicatic & CHON/organic \\
   \hline
   Size classes & 6 & 6\\
   Size ranges (m) & 10$^{-6}$ - 10$^{-1}$ & 10$^{-6}$ - 10$^{-1}$ \\
   Density (kg $\;$ m $^{-3}$) & 2.5 $\cdot$ 10$^{3}$ & 1.0 $\cdot$ 10$^{3}$\\
   Conductivity (grain) (K $\;$ kg$^{-1} \;$ m$^{-3}$) & 3 & 0.25 \\
   Conductivity (mantle) (K $\;$ kg$^{-1} \;$ m$^{-3}$) & 0.01 & 0.2\\
 \hline
\end{tabular}
\end{table}

\noindent For the grains of the first distribution a thermal conductivity value of 3 W~K$^{-1}$m$^{-1}$, typical of silicates, has
been adopted, while for the CHON-dominated grains a value of 0.25 W~K$^{-1}$m$^{-1}$, typical of organic substances
thought to be analogues of cometary organic material (K\"omle et al. \cite{koemle}), has been preferred. Average
density is 2500 kg~m$^{-3}$ in the first distribution, and 1000 kg~m$^{-3}$ in the second one, accordingly with Giotto
findings. \\
The size distributions of dust particles in the nucleus could be different from those measured in the coma: observed
particles are the smaller ones, and larger particles can be seen only by radar (Harmon et al. \cite{harmon}). The observed
exponential distributions
could be the results of fragmentation processes in the inner coma. A way to define an initial size distribution in the
nucleus could be to suppose that the grains are the result of the coagulation processes acting in the protosolar nebula:
following this idea, we attributed to both kind of
grains a gaussian distribution taken from a paper by Coradini et al. (\cite{coradini}) on the accretion of grains in the solar
nebula. \\
It remains to define a value for the conductivity of a silicate-dominated crust and for the conductivity of a crust with
a high abundance of organic refractory material. It was found, during KOSI experiments (Gr\"un et al. \cite{gruen}), that
thermal conductivity of dust layers composed of silicatic grains is very low
($\sim  10^{-2}$- 10$^{-3}$ W~K$^{-1}$m$^{-1}$): even a thin layer of dust causes a large temperature gradient. Following
this result, we attributed a value of 0.01 W~K$^{-1}$~m$^{-1}$ to the thermal conductivity of a crust composed only of
silicatic particles.  In the case of the "organic" distribution, lacking for direct measurements, we deduced the physical
properties from the results of the laboratory experiments conducted at the Space Research Institute of Graz
(K\"omle et al. \cite{koemle}). The thermal evolution of an analogue of cometary material composed of a mixture of minerals, ices
and hydrocarbons has been studied, obtaining a sort of cohesive crust composed of minerals glued together by organics.
This kind of crust has a thermal conductivity higher than that of  silicatic crust, and also has a greater cohesive
strength: an admixture with organic material raises crust conductivity of one order of magnitude or more with respect to a
crust composed of silicatic grains. So, our "organic" crust is characterized by a higher thermal conductivity than the
"silicatic" one: we attributed to a crust mainly composed of organic particles a value of 0.2 W~K$^{-1}$m$^{-1}$.\\
At the beginning of computations the dust particles are embedded in the porous ice. When the ice sublimates the embedded
particles become free, in a number proportional to the amount of sublimated ice, and can undergo the drag exerted by the
gas flux. Average pore size is increasing due to the ice sublimation: a "pore widening factor" proportional to the amount
 of sublimated ice is locally applied to the pore average size. In this way we are taking into account that a sublimating
ice produces holes and cavities, and as a consequence free particles can move towards the surface (particles can move and
be blown off only if their size is less than the local average pore size!). Free particles that are found close to the
surface can move toward the surface and be blown off or accumulate to form a dust mantle. Free particles are moving under
the opposite effects of gravity and gas flux; only grains smaller than the pore average size can move toward the surface
in the pore system. When the ice is condensing, a proportional
amount of free particles become embedded and the pores becomes smaller (porosity decreases).\\
To determine how many particles can be blown off to contribute to dust flux, and how many can instead be accumulated on
the surface, the different forces acting on the single grain are compared, obtaining for each distribution a critical
 radius, depending on grain density. The critical radius represents the radius of the largest particle that can leave the
 surface of the comet:

\begin{equation}
\label{rstar}
a^* = \frac{3}{4} \frac{\Phi_{H_{2}O} \cdot V_{H_{2}O}}
{\rho_{dust}[G \frac{M_n}{R_n^2} - R_n \omega^2 cos^2\:\theta]}
\hspace{1cm} [m]
\end{equation}

\noindent where $G$ is the gravitational constant, $M_n$ the mass of the comet, $R_n$ its radius, $\omega$ its angular
rotation velocity, $\rho_{dust}$ the dust grain density, $\Phi_{H_{2}O}$ is the gas flux,
 and $V_{H_{2}O}$ the velocity of the gas flux. In the Eq.~(\ref{rstar}) the numerator
represents the lifting force exerted by the outflowing gases, and the denominator represents the gravitational attraction
corrected by the centrifugal force. At each time step the number of free dust particles for each size class and the value
of the critical radius are computed. All the free particles with radius $a~<~a^*$ are considered ejected (blown off) and
contribute to the dust flux, while those with $a~\geq~a^*$ accumulate on the surface. This refractory surface layer (dust mantle) is not necessarily stable, and can be later on blown off if the gas flux is strong enough. It could also
increase in thickness if more and more particles are not able to leave the surface, substantially reducing and then
quenching the gas flux. Dust mantle compaction due to a volume reduction when ice is sublimated is taken into account.
If the dust mantle is not too thick the gas flux from the interior is able to entrain a dust flux, so it is possible to
have a dust flux even when there is no ice left on the surface.\\
In this way, even an originally homogeneous body becomes quickly highly inhomogeneous: composition, density and porosity
are changing with depth.

\subsection{Thermal modeling of P/Read}

A thermal model depends not only on the assumptions of the model itself, but also on the values attributed to the initial
parameters. These parameters describe the orbit, the initial state of the body and the properties of the matter of which
it is composed. They are defined or derived, when possible, from observation and laboratory experiments. A given set of
these parameters define a "Case". By changing a small number of these parameters, and keeping fixed all the rest, we can
build different Cases that are the subject of the simulation.\\
To apply the above described model to a real case we need to define the characteristics of the body that we want to
simulate, assuming an initial structure and composition. As for the real case, we chose to simulate the behaviour of
P/2005 U1 (Read). The results of this simulation can be applied also to the other MBCs known until now, because orbits and
sizes are reasonably similar.  \\
As written before, we start from the assumption that the model body has a cometary nature, that is it is composed of ice
and dust grains. A further assumption is being made that the surface of the body is covered by a devolatilized mantle,
and that a recently happened impact has opened a crater in this mantles. As a consequence, the heat from the Sun is now
able to reach ice-rich layers, so triggering a cometary-like activity (sublimating gas entraining dust particles).  At
the beginning of the simulation, the immediate effects of the impact (release of heat) have already ceased.\\
As for the composition and physical properties, it is difficult to make safe assumptions. We have no reason to think,
anyway, that these bodies are absolutely similar to the classical comets (Jupiter-family, Halley-family and Long Period
comets), because their origin and evolution history is different. They formed closer to the Sun than the comets now
stocked in the Kuiper Belt or Oort cloud. During the Solar System formation period materials from different distances from
the Sun were mixing up and the snow line could have been closer to the Sun than nowadays: this can explain the presence of
dust-ice bodies orbiting what is now the Main Belt. Collision evolution also played a major role in the past of these
bodies, probable survivors of the catastrophic event that gave origin to Themis family. Hardly the ice now remaining in
MBCs was formed in the amorphous state, because this requires extremely low temperatures (much lower than 50 K). For
similar reasons, it is extremely improbable that this ice could still be mixed with other ices more volatile than water
 that, even if they existed, should have been disappeared long time ago. We will make the assumption that the model body
contains only H$_{2}$O crystalline ice. \\
Some of the values given to the initial parameters of the model are listed in the table \ref{tab3}. The properties
attributed to the dust distribution are listed in the first column of the table \ref{tab2}. The values given to orbital
parameters (semimajor axis and eccentricity) and to the radius of the model body are derived from observation, and refers
to P/2005 U1 (Read). The values given to surface albedo, emissivity, and pore radius can be considered typical of comets,
 and are largely used in comet modeling (Huebner et al. \cite {huebner}). The spin period is not known, so we are using
a value (10 hours) typical of many small bodies.  The value for dust/ice ratio we are using in this simulation is 3,
higher than the one usually assumed for comets (1), as a way to take into account the origin of the comet in a zone of
the Solar System less rich in volatiles. The porosity is 0.4. The initial temperature, 130 K, has been defined taking into
 account the orbit of the body. \\

\begin{table}
\caption{I. Initial value of the parameters defining the model body.} \label{tab3}
\centering
\begin{tabular}{l c}
\hline \hline
   a & 3.165  \\
   e & 0.253  \\
   Dust/ice & 3  \\
   Initial temperature (K) & 130  \\
   Surface albedo & 0.04    \\
   Surface emissivity & 0.96    \\
   Spin period (hours) & 10  \\
   Initial pore radius (m) & 10$^{-4}$  \\
   Initial porosity & 0.4  \\
   Initial radius (km) & 0.3 \\
   \hline
\end{tabular}
\end{table}

\noindent The dust component is described as a distribution of 6 size classes with similar density and thermal conductivity
(see table \ref{tab2}). \\
 The model was run in the fixed-latitude approximation. The reasons for this choice are that the rotation period is
not known, and the variations of gas and dust production with respect to the daily hour are irrelevant for the present
analysis. In this way, gas and dust production rates depend only on the latitude and on the position along the orbit. \\

\subsection{Results: exposed ice}
\begin{figure}
\centering
\resizebox{\hsize}{!}{\includegraphics[width=3cm]{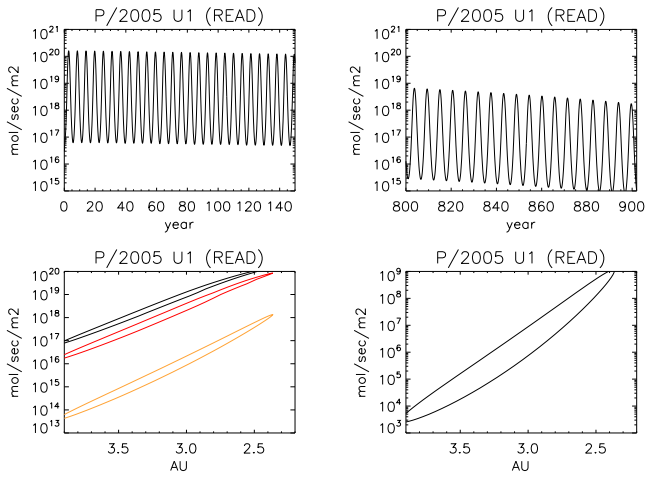}}
\caption{Case 1 - Water flux along the orbit for a P/Read-like body on which an impact has exposed ice-rich layers. Upper
left panel: impact at the  equator, water flux with respect to time, for the first 140 years after the impact; upper right
 panel: impact at the  equator, water flux with respect to time for a period of one hundred years immediately before the
disintegration of the available icy material; bottom left panel: impact at the  equator (upper curve) and at $30 \degr$
(lower curve), water flux with respect to heliocentric distance. The fluxes are shown 10 years after the impact; bottom
 right panel: impact at $85 \degr$, water flux with respect to heliocentric distance. The flux is shown 10 years after the
 impact.}
\label{fig1}
\end{figure}

\begin{figure}
\centering
\resizebox{\hsize}{!}{\includegraphics[width=3cm]{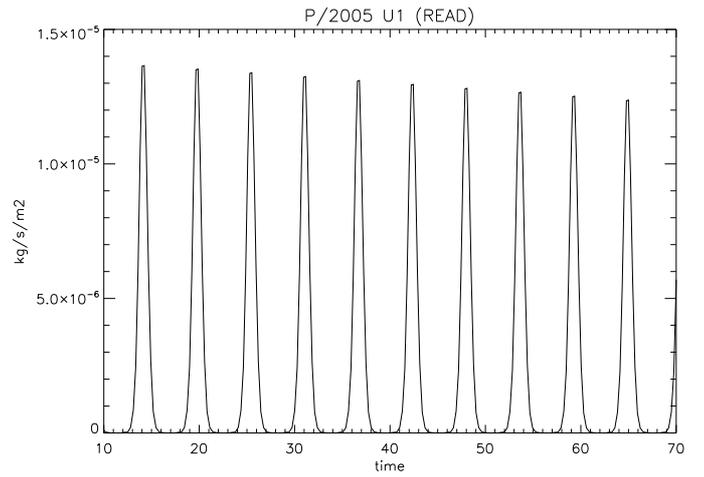}}
\caption{Case 1 - Impact at the equator, dust flux with respect to time for a period of 60 years following the impact.}
\label{fig2}
\end{figure}

\begin{figure}
\centering
\resizebox{\hsize}{!}{\includegraphics[width=3cm]{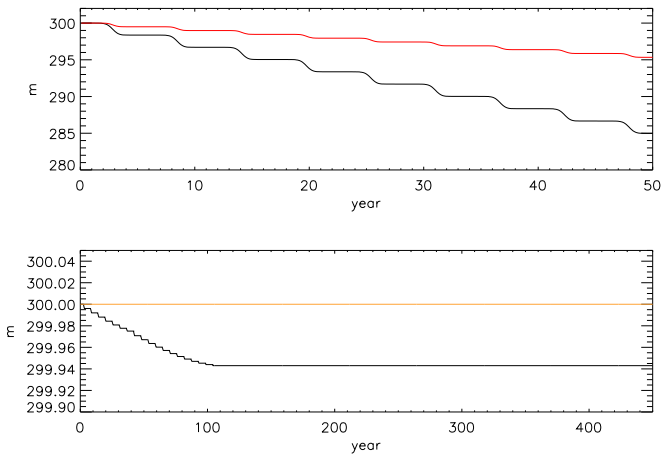}}
\caption{Case 1 - Stratigraphy showing the erosion of the surface with respect to time. Upper panel: impact at the equator
 (lower black curve) and at $30 \degr$(red curve); lower panel: impact at 60 \degr (lower black curve) and 85 \degr
(orange curve). At the beginning of computations, the surface has always a radius of 300 m. Depending on the latitude
of the impact, the radius shrinks with respect to time. }
\label{fig3}
\end{figure}

\begin{figure}
\centering
\resizebox{\hsize}{!}{\includegraphics[width=3cm]{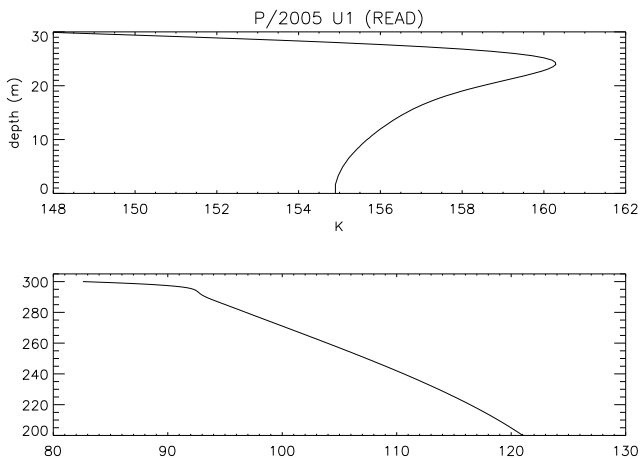}}
\caption{Case 1 - Internal temperature profile for a P/Read-like body at the aphelion, 900 years after the impact; impact
 at the equator (panel a) and at $85 \degr$ (panel b.}
\label{fig4}
\end{figure}

Case  1 has been ran assuming that the impact has directly exposed to the sun heating an icy zone located at given
cometocentric latitude. The model is ran for 4 different latitudes of the impact crater: equator (sub-solar point),
$30 \degr$, $60 \degr$  and $85 \degr$. The dust component is represented by one distribution, the silicatic one (first
column of table \ref{tab2}). With these parameters, bulk thermal  conductivity along the whole evolution has an average
value of 0.07 K $\;$ kg$^{-1} \;$ m$^{-3}$; average density values for the matter composing the model body are of the
order of 586 kg $\;$ m$^{-3}$.\\
The results can be seen in the figures \ref{fig1}, \ref{fig2}, \ref{fig3} and \ref{fig4}, showing, respectively, gas and
dust fluxes, stratigraphy and internal temperature profile. After a time of almost 900 years, the ice material exposed by
the impact at the subsolar point is completely consumed, and gas flux, that has been continuously reducing (\ref{fig1},
upper panels), is completely quenched. At all the latitudes, most of the dusty material is left behind: not all the dust
grains freed from the ice are going to contribute to the dust flux. At higher latitudes the situation is different, as can
 be seen in the bottom left panel of figure \ref{fig1}, showing the gas flux with respect to heliocentric distance for
impacts at the equator, $30 \deg$ and at $60 \deg$. In the bottom right panel of the same figure the water gas flux with
respect to heliocentric distance is shown for an impact at a latitude of $85 \deg$. In this case, the flux is extremely
reduced. Bottom panels show the gas flux ten years after the impact exposing the ice-rich layers.\\
  The dust flux with respect to time for an impact at the equator on a period of 60 years, starting 10 years after the
impact, is shown in figure \ref{fig2}. The reduction of the dust flux from orbit to orbit can be clearly seen. For impacts
 at higher latitudes, the dust flux is correspondingly reduced, similarly to what happens to the gas flux, and is almost
null when the impact crater is supposed to be located close to the pole.

The erosion level of the surface of the crater varies with the latitude of the impact and is proportional to the intensity
 of dust and gas emissions, as can be seen in figure \ref{fig3}, showing the stratigraphy (surface radius, in meters, with
 respect to time in years) for a period of 50 years after the impact: it is almost null close to the poles (lower panel),
where the solar input is minimum, while at the sub-solar point the available ice is consumed in less than one thousand
years (almost two meters of surface lost per orbit, upper panel). \\
The temperature profiles shown in figure \ref{fig4} are photographing the situation at the aphelion, after almost 9
 hundreds years of revolutions around the Sun. At the equator only 30 meters of active material remain (panel a), while at
85 $\deg$ of latitude (panel b) not even 1 centimeter of surface has been lost. Surface and interior temperatures give
 an explanation for this difference: at temperatures equal or less than 145 K, as is the case at latitudes close to the
 pole, water ice can be considered stable against sublimation. At equator, in the 30 meters of icy material remaining
after 900 years of activity, interior temperatures are much higher, always well above the sublimation limit. The first
 5 meters under the surface are under the influence of the orbital heat wave. In the deeper layers the temperature slowly
 decreases until it reaches 155 K, 25 K higher than the temperature at the time of the impact (130 K). \\

\subsection{Results: mantle-covered ice}

In order to trigger some activity, it is not necessary that the impact exposes directly icy-rich layers to the action of
the Sun. The result of a shallow impact in a devolatilized mantle could be excavate a crater, without directly exposing
ice-rich layers, but leaving only a thinner dust layer covering regions that had remained safely buried for millions of
years. In this case it could be possible not only activate the body (as a consequence of the impact itself) but also
 generate a stable or periodic activity, although fainter than in the situation described in the preceding section. It is
well known that ice sublimation can be possible from under a mantled surface (see, for example, Capria et al.
\cite{capria2}), providing that ice-rich layers are reasonably close to the surface and the mantle itself is somewhat
porous. This phenomenon has been observed, for example, on the comet 9P/Tempel 1, target of Deep Impact. The few small
patches of icy surface visible in the returned images are not enough to account for the activity of the comet, if
sublimation from the mantled zones is not considered (Belton and Melosh \cite{belton}). \\
To simulate this kind of activity, 4 Cases have been ran with a different assumption than in the preceding section: the
 bottom of the crater is still covered by a dust layer that has survived the impact. In this case, the duration and
intensity of the activity depend strongly on the mantle thickness and physical properties. It would be impossible to run
 Cases exploring all the different combination of input parameters having a direct influence on the results. Two different
 thicknesses (0.1 and 0.5 m) and two extremely different compositions (silicatic and organic/CHON) have been considered
for the surviving dust layer, obtaining 4 Cases (see table \ref{tab4}. The thickness of the mantle before the impact
should have been much larger: in order to keep sealed volatile materials on an extremely long timescale, tens of meters of
 insulating materials are necessary (see also Schorghofer \cite{schorghofer}). The surviving dust layer thicknesses to be
tested have been chosen somewhat arbitrarily, on the basis of our preceding experience on comet nucleus modeling. The
 results of these Cases can be easily extrapolated to different thicknesses of the surviving dust layer. \\
The aim of testing two different dust compositions is instead to show how the characteristics of the activity would depend
 on the physical properties of the involved materials: we are aware that pure silicatic or pure organic dust are somewhat
 unrealistic. The different compositions correspond in the code to two dust distributions with the characteristics listed
 in table \ref{tab2}. The remaining input parameters are the same as the ones listed in \ref{tab3}. The impact point is
 always assumed at the equator (sub-solar point).\\

\begin{table*}
 \begin{minipage}[t]{\columnwidth}
\caption{Input parameters characterizing the surviving dust layer at the point of the impact in the 5 Cases} \label{tab4}
\centering
\begin{tabular}{l c c c c c}
\hline \hline
    Case      & 1 & 2 & 3 & 4 & 5\\
   Dust distribution & silicatic & silicatic & silicatic & organic & organic \\
   Thickness of the dust layer at impact location (m) & 0 & 0.1 & 0.5 & 0.1 & 0.5 \\
     \hline
\end{tabular}
\end{minipage}
\end{table*}

\begin{figure}
\centering
\resizebox{\hsize}{!}{\includegraphics[width=3cm]{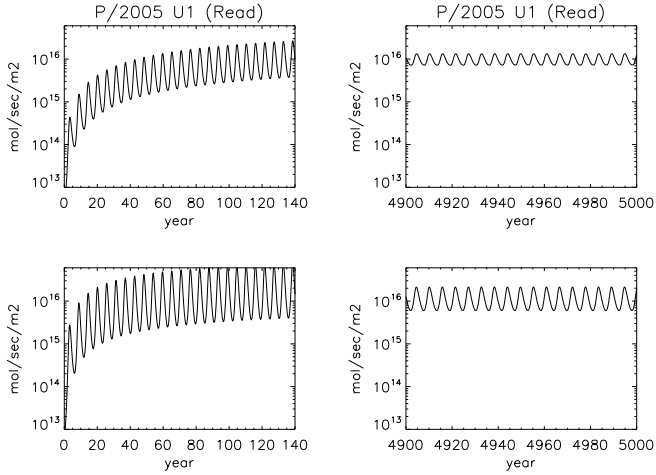}}
\caption{Cases 2 and 4 - Water fluxes with respect to time, at the beginning and at the end of computation. The thickness
 of the surviving dust layer is 0.1 m. Upper panels: Case 2 (silicatic dust). Bottom panels: Case 4 (organic dust).}
\label{fig5}
\end{figure}

\begin{figure}
\centering
\resizebox{\hsize}{!}{\includegraphics[width=3cm]{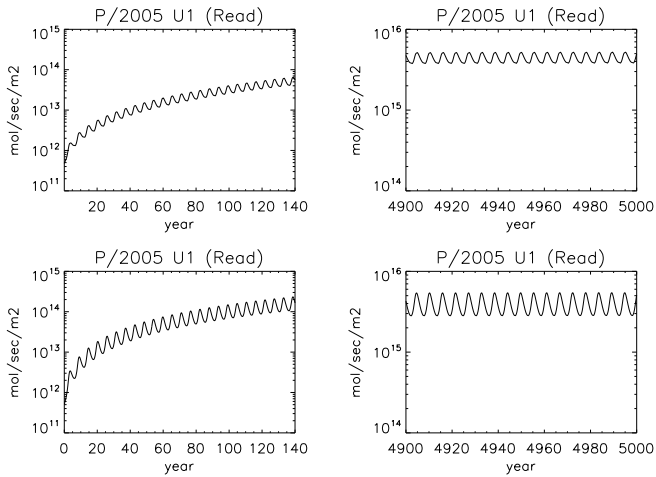}}
\caption{Cases 3 and 5 - Water fluxes with respect to time, at the beginning and at the end of computation. The thickness
 of the surviving dust layer is 0.5 m. Upper panels: Case 3 (silicatic dust). Bottom panels: Case 5 (organic dust).}
\label{fig6}
\end{figure}

\begin{figure}
\centering
\resizebox{\hsize}{!}{\includegraphics[width=3cm]{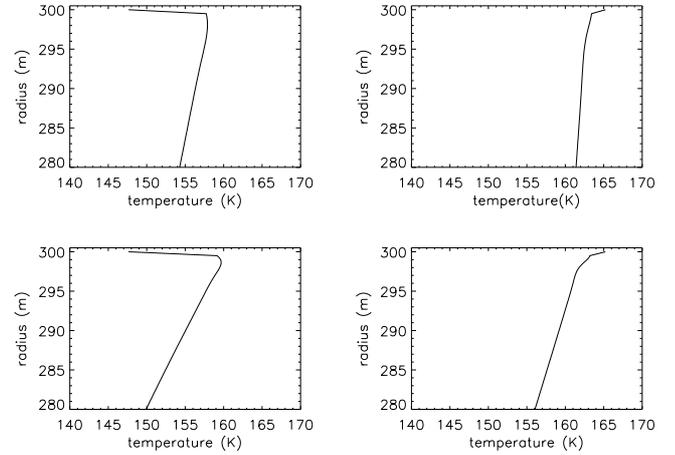}}
\caption{Cases 3 and 5 - Temperature profiles at equator. Left upper panel: Case 3, 1700 years after the impact, aphelion.
Right upper panel: Case 3, 5000 years after the impact, 3.16 AU. Left bottom panel: Case 5, 1700 years after the impact,
aphelion. Right bottom panel: Case 5, 5000 years after the impact, 3.16 AU. }
\label{fig7}
\end{figure}

\noindent
Figures \ref{fig5}, \ref{fig6}, and \ref{fig7}show the result of these 4 Cases. In fig. \ref{fig5} the gas flux
is shown for a body with a 0.1 m thick surviving mantle layer, assuming a silicatic composition (upper panels, Case 2) and
an organic composition (bottom   panels, Case 4). The left panels show the evolution of the water flux in the years
following the impact, while the right panels show the flux after almost 500 years after the impact. A similar figure
(\ref{fig6}) show the result of Cases 3 and 5, where the thickness of the surviving mantle is 0.5 meters.\\
        A quite obvious result is that the water ice is sublimating with a rate much lower than in Case 1, depending on
the thickness of the surviving dust layer: between Cases 2, 4 (0.1 m thick layer) and Cases 3, 5 (0.5 m thick layer) there
are orders of magnitude of difference. Less obvious are the long-term evolution, and the dependence of the activity on the
 properties of the material forming the dust layer.  When the computations begin, the interior temperature of the body is
homogeneous and equal to the given initial temperature, 130 K in this case. With the passing of the years, the interior
 temperature slowly increases, and this corresponds to an increase in the sublimation rate. The process goes on till the
 interior temperature stabilizes, along with the sublimation rate (right panels in figures \ref{fig5} and \ref{fig6}.
Surface temperatures, mainly determined by solar input, are spanning the range from 148 K to 191 K in all the four Cases. \\
        Due to the action of two competing processes, the thickness of the surviving dust layer changes, even by few
millimeters, in the time after the impact. After 5000 years, in Cases 2 and 4 the thickness of the mantle has increased,
 at the bottom, by 15 centimeters and almost 12 centimeters respectively. This increase is the result of the slow
devolatilization of the icy-reach layers adjacent to the bottom of the mantle. During the same time, a faint dust flux has
 eroded the surface by 2 and 3 millimeters in the two Cases. As for the Cases in which the initial thickness of the mantle
 was higher, after 5000 years of evolution less than 5 centimeter of devolatilized material has been added at the bottom
 of the crust, while the surface erosion is less than 2 millimeters in Case 5 and nothing in Case 3.\\
        The different values attributed to the thermal conductivity and density of the refractory component of the model
 body (see table \ref{tab2}) explain the differences in the activity between the Cases 2,3 and 4, 5. Due to the higher
 thermal conductivity of silicatic grains, in Cases 2 and 4 higher internal temperatures are attained with respect to
 Cases 4 and 5. Figure \ref{fig7} shows the temperature profiles of Cases 3 and 5, at the beginning (left panels) and
towards the end of computation (right panels). The strong temperature gradient between the surface and the sublimation
front (where water ice, mixed with dust, still survives), is visible, along with the difference in surface temperature
between the aphelion, at 3.96 AU, and a point on the orbit, 3.16, closer to perihelion (2.36 AU). \\

 At a certain point, the increase in the internal temperature ceases because an equilibrium temperature has been reached.
 This process has a different duration depending on the physical properties of the materials forming the body. It can be
 noticed that when the computation stops, 5000 years after the impact, the model bodies are still active, and it is
reasonable to assume that they will continue to be active for a very long time. As for the thickness of the surviving
dust layer, there are two competitive processes at work: dust emission from the surface (and the closest layers), and
progressive devolatilization of layers close to the bottom of the mantle. The second process is the winner, and so the
 thickness of the mantle is slowly growing. As a consequence, the gas flux is slowly reducing. In order to study how much
time would the body remain active, much longer, and heavy, computations would be required. This will be the subject of a
future work.

\subsection{Comparison with observation}

The results of the theoretical models should now be compared with observations. Only in this way would it be possible to
judge which ones, between the different Cases presented in the above sections, are closer to reality. Unfortunately, what
we have from observation about P/Read is not enough to make possible such a comparison. From the paper by Hsieh et al.
(/\cite{hsieh1}), who observed this object for almost three months in 2005 and one night in 2007, P/Read has been always
found active in 2005, while in 2007 it was too faint to permit an estimation of the activity level. It must be noted that
 P/Read in July 2005 passed at its perihelion, and the aphelion passage was in May 2008, a few months after
the observation of 2007. In 2005, coma and tail were clearly visible. Monte Carlo numerical simulation of dust emission
indicates that coma and tail were optically dominated by dust grains greater than 10 micron. Mass-loss rate has been
 estimated to be of the order of $0.2$ kg s$^{-1}$. Dust emission probably began at least few months before the discovery
of the activity (October 2005). \\
This estimation of dust emission rate cannot be directly compared with the results of the 5 Cases shown before, because we
 have no idea if the activity is confined at the bottom of a crater or if it is instead is widespread on large areas of
 the surface. In order to give some numbers, let us make the assumption that only the bottom of a recently excavated
crater at the subsolar point is the source of the observed activity (the walls would not contain volatile material). The
 surface area of P/Read is approximately $1.1 x 10^{6} $m$^{2}$. If we assume a dust production rate, at the equator, of
 $8.8 x 10^{-6} $kg s$^{-1} $m$^{-2}$ (figure \ref{fig2}), in order to obtain the observed dust emission rate we need an
 active area of approximately $2.3 x 10^{4} $m$^{2}$. The evaluation, in turn, gives a very rough estimation of the depth
of the crater, if we assume a bowl-like shape. The implied crater diameter, for a planar surface, is $\sim 170$~m.
Assuming a depth-to-diameter ratio of 0.2 (Melosh \cite{melosh}), the corresponding crater depth is $\sim 34$~m\.
This number cannot be interpreted as a precise measure of the thickness of the original mantle of P/Read (could have been
thinner at the impact point), but is only useful to give an idea of the orders of magnitude involved. This rough
computation is performed for a crater close to the subsolar point: in case of an impact at a higher latitude, all the
numbers should be scaled correspondingly.\\
The crater size estimate given above is a lower limit. It would apply only for inert mantle thickness much lower that the
estimated crater depth. However, we expect that MBCs have mantles thickness  of the order of tens of meters because
thinner mantles would have not have been able to prevent the complete  sublimation of volatiles for MBC-sized objects.
 For instance, in the case that a thickness of half of excavated depth consist of inert mantle, the same active area is
obtained for a crater diameter about $\sqrt(2)$ times larger.\\
Looking to the plots in figures \ref{fig1}, \ref{fig2}, \ref{fig5} and \ref{fig6}, the hypothesis of an impact leaving a
mantled bottom appears much less compatible with observation, unless we consider a surviving dust layer much thinner that
 in the Cases 2, 3, 4 and 5: the activity level is so faint that it would be hardly detected from ground.

\section{How frequent are the impacts?}

In order to address if cratering events are a viable process to explain the observed activity and number of MBCs, it is
important to estimate not only the size of the crater required to sustain the observed activity, but also the frequency of
 its formation.\\
  The estimate of the projectile size needed to form a crater of 170~m has been performed considering two scaling laws,
 namely for porous material and cohesive soils (Holsapple and Housen \cite{holsapple}). The material strength has
 been set to $2x10^4$ and $10^7~$dyne/cm$^2$. We assumed a target and projectile density of 1000 kg/m$^3$ and 2600
kg/m$^3$  respectively. Moreover, the computed P/Read intrinsic average impact probability within the Main Belt is about
3.48x10$^{18}$~km$^{2}$~yr${^-1}$ and the average impact velocity is about 3.7~km s$^{-1}$. These values are computed
according to the Bottke et al. (\cite{bottke}) Main Belt population, and following the procedure described in the
work of Marchi et al. (\cite{marchi}). The derived projectile sizes are $\sim 8, 18.5$~m for porous and cohesive
material, respectively.\\
The computed time for having a collision with a projectile $\ge 8$~m on a body with the size of P/Read (0.6~km) is
5.1x10$^7$~yr. The estimated total number of objects larger than 0.6~km is $2.9x10^6$,
therefore such a collision is expected to occur on average every $\sim17$~yr everywhere in the Main Belt. This time become
$\sim 147$~yr for projectiles $\ge$ 18.5 ~m.\\
Of course these numbers have to be taken as a rough indication, given the uncertainties in the Main Belt size distribution
 at such small size and in the target material parameters. Note that the
timescales above also depend on the target size. For a body having the size of Elst-Pizarro (5~km), it is estimated that
the same impacts occur every $\sim 10$ and $\sim 100$~yr, respectively  for projectiles of 8 and 18.5~m.

\section{Discussion}

When running a simulation, it would be impossible to test all the possible combination of input parameters. What is
usually done is, once found a meaningful combination,  to keep fixed values for most of the parameters and vary only one
or two at a time, in order to test the influence of the changes on the results. The variable parameters are the ones
considered to be most influential on the properties or phenomena that are the object of the simulation. In this paper, an
initial  composition for P/2005 U1 (Read) has been assumed, and we have varied, to build the 5 Cases that have been
studied, the parameters defining the thickness and composition of the dust mantle. As for the possible effects of changing
 the value attributed to the remaining initial parameters, it is here discussed  on the basis of our experience.\\
 A lower value of dust/ice, (for example, a value of 1, usually attributed to classical comets) would give as a result a
 lower dust flux, and consequently a slower erosion of the surface. Changing the thermal conductivity, porosity, density,
 pore radius would have the effect of slowing or quickening the evolution of the model body. The thickness of the
surviving mantle in the Cases 2, 3, 4 and 5 has been chosen in a quite arbitrary way. It is clear, anyway, that a larger
 thickness  would tend to completely quench any possible activity. The mantle thickness necessary to completely block the
 underneath sublimation strongly depends on the average porosity, thermal conductivity and the other physical properties
of the dust layer.\\
  The issue of the initial temperature attributed to the model body deserves some more discussion. Surface temperature is
 mainly determined by the solar input, and is spanning, along the orbit followed by the model body, the range 147 - 190 K
 (subsolar point, fast rotator approximation).  Obviously, nobody knows what really is the average internal temperature
(below diurnal and seasonal skin depth) of such a body, but the assumption has been made that it should be around 130 K,
 a value at which water ice can survive. When the insulating mantle is disrupted, the internal temperature is steadily
rising till an equilibrium value is reached. A different value of the initial temperature would result in a slowing or
quickening of this process. \\
 In this paper the possible effect of a single impact has been simulated. During the long time spent orbiting around the
 Sun, many  small impactors hit the surface of bodies in the Main Belt. Even if these impacts are not able to disrupt the
 insulating mantle, it is probable that they are producing some changes on the surface. The effect of the impact itself,
 and of the released heat, could be that, for example, of locally compacting a porous surface mantle and locally changing
 its properties. \\
  The results obtained for P/ 2005 U1 (Read) can be easily applied to the other MBCs known until now, because they are all
 revolving on similar orbits, with the only possible exception of P/2008 R1 (Garradd), that is coming closer to the Sun at
 perihelion. For this object, solar input is slightly higher and the results obtained for P/2005 U1 (Read) should be
consequently scaled.\\
From the results of these simulation (in particular of Cases 2 to 4) , it is also possible to understand how icy material
 buried deep enough under an insulating mantle could have survived for such a long span of time. A thorough analysis of
this process will be the subject of a future paper. As for the origin of these ice-rich layers still surviving in bodies
 orbiting well inside the snow line border, it must probably be sought in the chaotic processes that were shaping the
solar system at the beginning of its history, bringing bodies away from their formation places.

  \section{Summary and conclusions}

Assuming that the newly discovered Main Belt Comets are icy bodies on which a stable cometary activity has been induced
by a recent impact that excavated a crater in a dust mantle, two different scenarios have been simulated. In the first
one, it has been assumed that  ice-rich layers, remained buried under an insulating dust layer for a very long period of
 time, are now directly exposed to Sun heating. In the second one, it has been assumed that the bottom of the newly
excavated crater is still covered by a layer of dust much thinner than before the impact, so that the solar input is now
able to reach ice-rich layers. The characteristics attributed to the model body are those of P/2005 U1 (Read). It has
 been demonstrated that, given the existence of buried ice, the activity observed in the MBCs can be explained as cometary
 activity, that means dust emission sustained by ice sublimation.  \\
       In the first scenario, a stable cometary activity can be obtained, compatible with observations. Making a rough
estimate, a crater at the subsolar point with a diameter of 170 meters would be compatible with the observed dust
production rate. The activity would be stable, but the water ice is quickly consumed, depending on the cometocentric
latitude of the impact point. At the subsolar point, the surface erosion could reach two meters per orbit. When, as
 assumed in the second scenario, a dust layer still remains, covering the bottom of  the crater, a stable activity is
still possible, providing that the heat of the Sun is able to reach ice-reach layers underneath. The activity is much
fainter and long lasting than in the preceding case, depending on the latitude of the impact and the physical properties
 of the dust mantle. This kind of activity is probably not compatible with the activity observed on P/2005 U1 (Read).\\
        The computed time for having a collision with a projectile $\ge 8$~m on a body with the size of P/Read (0.6~km) is
 5.1x10$^7$~yr. The estimated total number of objects larger than 0.6~km is $2.9x10^6$,
therefore such a collision is expected to occur on average every $\sim17$~yr everywhere in the Main Belt. This time become
 $\sim 147$~yr for projectiles $\ge 18.5$~m. For a body having the size of Elst-Pizarro (5~km), it is estimated that the
 same impacts occur every $\sim 10$ and $\sim 100$~yr, respectively  for projectiles of 8 and 18.5~m. \\
                       There are no clues as how many other MBCs could exists in the Main Belt. Surely the great majority
 of bodies still keeping some ice in their interior are presently inactive, or their activity is so faint that it cannot
 be detected, at least from Earth-based observation.

\begin{acknowledgements}
      This work has been supported by ASI-INAF contract n. I/015/07/0 "Solar System Exploration".
\end{acknowledgements}
\bibliographystyle{plain}

\end{document}